\documentclass[%
preprint,
amsmath,amssymb,
aps,
]{revtex4-1}

\usepackage{graphicx}
\usepackage{dcolumn}
\usepackage{bm}
\usepackage{hyperref}
\usepackage{color}
\usepackage{setspace}
\usepackage{geometry}
\usepackage[sort&compress]{natbib}
\begin{document}
	\setlength{\baselineskip}{17pt}
	
	\title {Realization of the all-optical phase modulator, filter, splitter, and self-consistent logic gates based on assembled magneto-optical heterostructures}
	
	\author{Jie Xu$^{1,2}$, Yun You$^{3}$, Fengwen Kang$^{4}$, Sanshui Xiao$^5$, Lujun Hong$^6$, Yun Shen$^6$, Yamei Luo$^{1,2}$, and Kosmas L. Tsakmakidis$^{7}$}
	\affiliation{$^1$School of Medical Information and Engineering, Southwest Medical University, Luzhou 646000, China}
	\affiliation{$^2$Medical Engineering \& Medical Informatics Integration and Transformational Medicine of Luzhou Key Laboratory, Luzhou 646000, China}
	\affiliation{$^3$School of Science, East China Jiaotong University, Nanchang 330013, China}
	\affiliation{$^4$College of Materials Science and Engineering, Sichuan University, Chengdu 610065, China}
	\affiliation{$^5$DTU Fotonik, Department of Photonics Engineering, Technical University of Denmark, DK-2800 Kgs. Lyngby, Denmark}
	\affiliation{$^6$Institute of Space Science and Technology, Nanchang University, Nanchang 330031, China}
	\affiliation{$^7$Section of Condensed Matter Physics, Department of Physics, National and Kapodistrian University of Athens, Panepistimioupolis, GR-157 84 Athens, Greece}
	\affiliation{Corresponding authors: J. Xu (xujie011451@163.com), Y. Luo (luoluoeryan@126.com), and  K. L. Tsakmakidis (ktsakmakidis@phys.uoa.gr)}

	\begin{abstract}
		All-optical computing has recently emerged as a vibrant research field in response to the energy crisis and the growing demand for information processing. However, the efficiency of subwavelength-scale all-optical devices remains relatively low due to challenges such as back-scattering reflections and strict surface roughness. Furthermore, achieving multifunctionality through the reassembly of all-optical structures has thus far been rarely accomplished. One promising approach to address these issues is the utilization of one-way edge modes. In this study, we propose four types of deep-subwavelength ($\sim 10^{-2} \lambda_0$, where $\lambda_0$ is the wavelength in vacuum) all-optical functional devices: a phase modulator, a filter, a splitter, and logic gates. These devices are based on robust one-way modes but do not require an external magnetic field, which can allow for flexible assembly. In particular, we investigate a phase modulation range spanning from $-\pi$ to $\pi$, a perfect filter that divides the input port's one-way region into two output one-way regions with equal bandwidth, a multi-frequency splitter with an equal splitting ratio (e.g., 50/50), and self-consistent logic gates. We validate these theoretical findings through comprehensive full-wave numerical simulations. Our findings may find applications in minimal optical calculations and integrated optical circuits.
		
	\end{abstract}
	
	\maketitle

	\section{Introduction}
	Optical communication is known for its low loss \cite{tong:Sub}, parallel calculation capability \cite{feldmann:Parallel}, and high speed. As a result, all-optical devices are attracting increasing attention. Regular all-optical devices include, but are not limited to, optical filters, sensors, splitters, switches, frequency/phase/amplitude modulators, couplers, and amplifiers. Most of these devices rely on the electro-optic (EO) effect, thermo-optic (TO) effect, or nonlinear optical (NLO) effect. Among these optical effects, the linear EO effect has been extensively studied in fields such as optical sensing and optical frequency combs \cite{zhuang:electro} due to its ultrafast response \cite{vogler:Photonic}. The TO effect is commonly employed in the design of optical switches \cite{chen:Review}, while the NLO effect is primarily used for light control with light, enabling applications such as laser generation and optical switching \cite{ren:Tailorable}. However, all-optical devices based on EO, TO, or NLO suffer from high power consumption due to reflection induced by reciprocity/symmetry and fabrication errors.
	
	In recent years, there has been a significant focus on one-way edge modes due to their potential for breaking the diffraction limit (similar to SPPs) \cite{Xu:Ul,Tsakmakidis:Br,li:broadband}, robust processing/computing capabilities \cite{zangeneh:analogue}, and low loss. One common approach to achieve one-way edge modes is by utilizing photonic crystals (PhCs). The key lies in engineering a special surface that is bounded by materials with zero and nonzero Chern numbers. According to the bulk-edge correspondence \cite{tang:topological,ma:topological}, this ensures the existence of one-way edge mode(s) on this surface. The robustness of one-way modes in PhCs to imperfections or bends has been proposed theoretically and verified experimentally by several groups \cite{poo:Experimental,zhou:Photonic,Wang:To}. However, PhCs still pose challenges in terms of their relatively complex manufacturing process and the large device size compared to the wavelength of the guiding wave ($\lambda_0$). This poses a challenge for achieving subwavelength or even deep/ultra-subwavelength optical devices.
	
	Recently, our group has proposed several subwavelength magneto-optical (MO) one-way structures and discovered interesting phenomena/devices, such as unidirectional/bidirectional slow light \cite{Xu:Sl,Xu:Tr}, a perfect optical buffer with zero phase shift \cite{Zhou:Realization}, and all-optical logic gates (LGs) \cite{Xu:All}. The most significant advantage of the MO one-way structure is its extremely simple manufacturing process, requiring only two joint layers and not having strict requirements for surface roughness. Ultra-subwavelength all-optical devices can be achieved based on the MO structure, regardless of the nonlocal effect, which holds true in common scenarios, especially when the wavenumber is relatively small (e.g., $k<100 k_0$) \cite{Gangaraj:Do,Tsakmakidis:Topological}.
	
	Here, we propose the design of assembled MO structures and several interesting deep-subwavelength ($\sim 10^{-2}\lambda_0$) all-optical functional devices, including phase modulators, filters, splitters, and self-consistent logic gates (LGs). We study the propagation characteristics of surface magnetoplasmons (SMPs) \cite{Brion:Th} in theory, and demonstrate the low-loss properties of one-way SMPs through full-wave numerical simulations. Furthermore, we verify the functionality of these all-optical devices. The key to realizing the all-optical phase modulator, splitter, and self-consistent LGs lies in the utilization of one-way index-near-zero (INZ) modes. Unlike other INZ modes found in epsilon-near-zero (ENZ) \cite{Zhou:Br,niu:epsilon,Silveirinha:Tu}, mu-near-zero (MNZ) \cite{Marcos:Ne}, or epsilon-and-mu-near-zero (EMNZ) \cite{mahmoud:Wave} metamaterials, the INZ modes in our work exhibit a single propagation direction and can be easily adjusted within a continuous band.
	
	\section{Tunable INZ modes and all-optical phase modulator}
	\begin{figure}[hpt]
	\centering\includegraphics[width=6 in]{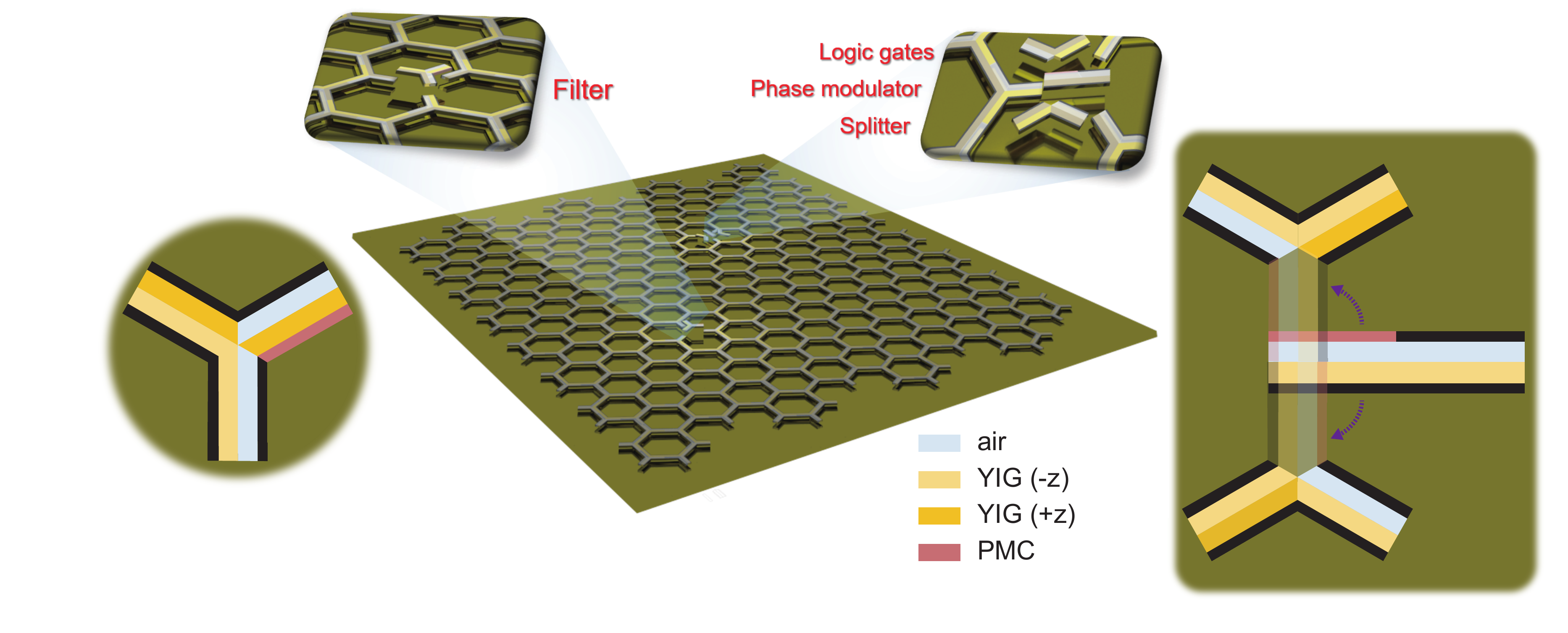}
	\caption{Assembled all-optical communication systems and functional devices utilizing YIG with remanence, along with PEC and PMC walls.}\label{Fig1}
	\end{figure}
	Yttrium iron garnet (YIG) stands out as one of the most captivating MO materials, finding widespread applications in optical communication. These applications encompass optical isolators \cite{fakhrul:Magneto,liang:Onchip}, band-pass filters \cite{popov:type}, high-efficiency optomagnonic micron-sized resonators \cite{almpanis:Spherical}, robust LGs \cite{Xu:All}, and more. Benefiting from the unique magnetization characteristics of YIG, residual magnetism persists even after removing the bias magnetic field. This property holds great potential for the realization of reliable all-optical communication, as depicted in Fig. 1. In this paper, we demonstrate that by utilizing YIG with remanence and PMC walls, a wide range of all-optical functional devices, including phase modulators, filters, splitters, and LGs, can be achieved. Importantly, these devices can be finely controlled and integrated into a system, as illustrated in Fig. 1.

	To investigate the impact of PMC walls on SMPs, we first construct four MO heterostructures surrounded by PEC and/or PMC walls \cite{balanis:advanced}. These structures are named as follows: the PEC-air-YIG-PEC (EDYE) structure, the PEC-air-YIG-PMC (EDYM) structure, the PMC-air-YIG-PEC (MDYE) structure, and the PMC-air-YIG-PMC (MDYM) structure, as illustrated in the inset pictures in Fig. 2. It should be noted that the YIG material with remanence are used in this study, which means that all the proposed functional devices in this work do not require precise control of an external magnetic field. Our theoretical analysis reveals the following forms for the dispersion relationship of these structures:

	\begin{equation}
		\alpha_r X_1 + \frac{\mu_{r2}}{\mu_{r1}}k-\alpha_d \mu_{vr} X_2=0
	\end{equation}
	where $X_1$ and $X_2$ are parameters that directly depend on the lower and upper boundary conditions of the structures, respectively. For the four structures, their values are as follows:
	
	\begin{equation}
		\mathrm{Lower \ boundary}
		\begin{aligned}
			& \begin{cases}
				&X_1 = \frac{1}{tanh(\alpha_r d_2)} \qquad (\mathrm{PEC}) \\	
				&X_1= \frac{\alpha_r \cdot tanh(\alpha_r d_2)-\frac{\mu_{r2}}{\mu_{r1}}k}{\alpha_r -\frac{\mu_{r2}}{\mu_{r1}} k\cdot tanh(\alpha_r d_2)} \qquad (\mathrm{PMC})	
			\end{cases} \\
		\end{aligned}
	\end{equation}
	\begin{equation}
		\mathrm{Upper \ boundary}
		\begin{aligned}
			& \begin{cases}
				&X_2 = -\frac{1}{tanh(\alpha_d d_1)} \qquad (\mathrm{PEC}) \\	
				&X_2= -tanh(\alpha_d d_1) \qquad (\mathrm{PMC})	
			\end{cases} \\
		\end{aligned}
	\end{equation}
	where $\mu_{r1}=1+i \frac{\nu \omega_r}{\omega (1+\nu^2)}$ and $\mu_{r2}=-\frac{\omega_r}{\omega(1+\nu^2)}$ are the nondiagonal elements of the tensor relative permeability of the YIG layers \cite{Shen:Large-area,Xu:Ul}. $\alpha_r$ and $\alpha_d$ represent the attenuation factors of the surface wave in the YIG layer and air, respectively \cite{Hong:Br}. $d_1$ and $d_2$ are the thicknesses of the air and YIG layers. $\omega_r$ is the characteristic circular frequency, which is denoted as $\omega_m$ when the YIG reaches saturation magnetization. In this paper, we assume $\omega_m=2\pi \times 5 \times 10^9$ rad/s and $\omega_r=2\pi \times 3.587 \times 10^9$ rad/s. It can be observed that as $k \to \pm \infty$, $X_1 \to 1$ and $X_2 \to -1$. Additionally, Eq. (1) tends to 
	\begin{equation}
		\begin{aligned}
			& \begin{cases}
				&1+\frac{\mu_{r2}}{\mu_{r1}}+\mu_{vr}=0 \iff \omega_{sp}^{+}=\omega_r \qquad (k \to +\infty)\\	
				&-1+\frac{\mu_{r2}}{\mu_{r1}}-\mu_{vr}=0 \iff \omega_{sp}^{-}=0.5\omega_r  \qquad (k \to -\infty)	
			\end{cases} \\
		\end{aligned}
	\end{equation}
	According to Equation (4), we can expect, at most, one asymptotic frequency for $k>0$, denoted as $\omega_{sp}^{+}$, and one asymptotic frequency for $k<0$, denoted as $\omega_{sp}^{-}$. Figure 2 illustrates the four dispersion curves of SMPs in these structures as $d_1=d_2=0.03\lambda_m$. Interestingly, when the lower PEC wall in the original EDYE structure is replaced with a PMC wall (see Fig. 2(b)), the one-way region abruptly changes from the higher region, $\omega_{sp}^{-}<\omega<\omega_{sp}^{+}$ (see Fig. 2(a)), to the lower region, $0<\omega<\omega_{sp}^{-}$, and simultaneously the propagation direction changes oppositely. Similarly, when the upper PEC wall in the EDYE structure is replaced with a PMC wall (Fig. 2(c)), the asymptotic frequencies and the one-way region remain the same, but the one-way SMPs consistently have positive values of k, i.e., $k>0$. Furthermore, when both the upper and lower PEC walls are replaced by PMC walls (Fig. 2(d)), the one-way region is completely disrupted due to the emergence of modes based on total internal reflection (TIR).
	
	In the inset of Fig. 2, we present simulation results with a setting of $\nu=10^{-3}\omega_m$, which shows good agreement with our theoretical analysis considering lossless materials ($\nu=0$). The corresponding transmission coefficients in these simulations are approximately $62.20\%$ (effective propagation length $L_{eff} \simeq 1.63\lambda_0$), $87.73\%$ ($L_{eff} \simeq 2.17\lambda_0$), and $86.82\%$ ($L_{eff} \simeq 5.06\lambda_0$). Thus, the proposed deep-subwavelength structures exhibit truly low loss.

	\begin{figure}[pt]
		\centering\includegraphics[width=4.5 in]{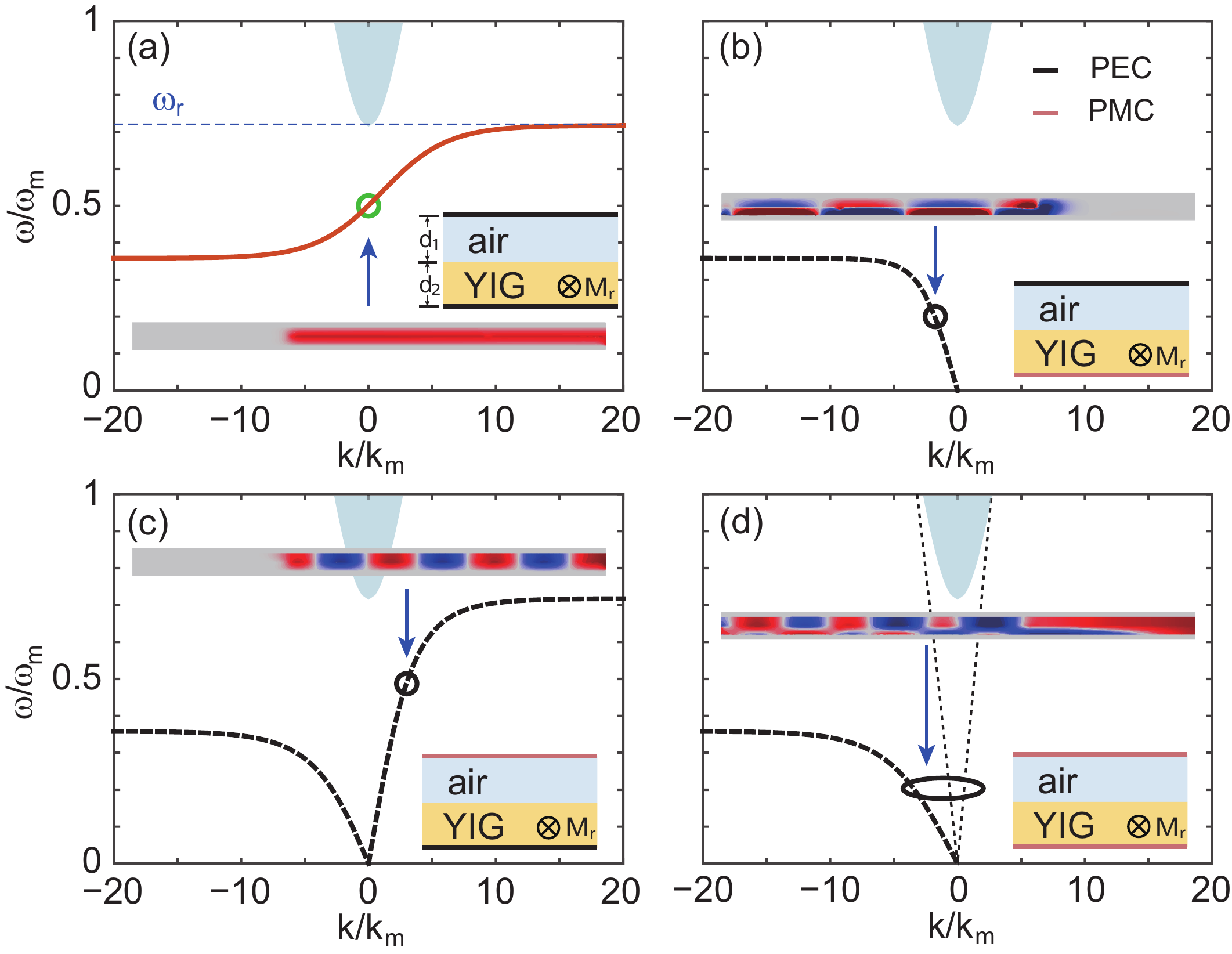}
		\caption{Dispersion diagrams of 'XDYX' structures where the first 'X' can be either (a, b) PEC or (c, d) PMC, while the second 'X' can also be either (a, c) PEC or (b, d) PMC. The red line and black dashed lines represent the dispersion curves of the SMPs. The cyan dashed area is the bulk zone of the bulk YIG. The other parameters are $d_1=d_2=0.03 \lambda_m$, $\omega_r=0.7174\omega_m$, and $\varepsilon_d=1$.}\label{Fig2}
	\end{figure}
	
	Furthermore, one of the most significant differences between Fig. 2(a) and Figs. 1(b,c) is the presence of an INZ mode, marked by a green circle in Fig. 2(a). In our previous work, we proposed a method to control the thickness parameters in order to adjust the INZ modes in magnetized-YIG systems. Similarly, in the current study, we can employ the same approach to achieve tunable INZ modes. The advantage of integrability in this system is undeniable compared to previous systems. 
	
	In Fig. 3, we investigate the impact of the thicknesses of YIG ($d_2$) and air ($d_1$) on the INZ modes in two cases. In the first case, we assume $d_1=d_2$ ($<0.1\lambda_m$, subwavelength). As shown in Figs. 2(a) and 2(b), when we increase $d_1$, the INZ frequency gradually decreases and remains close to $\omega=0.5\omega_m$. The red line in Fig. 3(a) represents a fitting curve with the following form:
	\begin{equation}
		\bar{\omega}=5.4848\cdot \bar{d_1}^2-0.09412 \cdot \bar{d_1} +0.5085
	\end{equation}
	where $\bar{\omega}=\omega/\omega_m$ and $\bar{d_1}=d_1/\lambda_m$ ($\lambda_m=2\pi c/\omega_m$). Equation (5) allows us to estimate the INZ frequency ($f_\mathrm{INZ}$) in the $d_1=d_2$ case. For example, when $d_1=0.045\lambda_m$, according to Eq. (5), the INZ frequency should be approximately $0.4932\omega_m$. The inset of Fig. 3(a) further confirms that when $d_1=d_2=0.045\lambda_m$, a point source with $f=0.4932f_m$ can excite SMPs with a sufficiently large effective wavenumber ($k_{eff}$) and exhibit no phase shift during propagation. In the second case, we assume $d_2=2 d_1$. Similar to the first case, we plot $f_\mathrm{INZ}$ as a function of $d_1$ in Fig. 3(c). The corresponding fitting equation is as follows:
	\begin{equation}
		\bar{\omega}=102.45\cdot \bar{d_1}^3-25.783\cdot \bar{d_1}^2+0.3716 \cdot \bar{d_1} +0.5828
	\end{equation}
	In comparison to the red line in Fig. 3(a), the red line in Fig. 3(c) exhibits a sharper slope, resulting in greater changes in $f_\mathrm{INZ}$ as $d_1$ increases. It is important to emphasize that, in theory, it is possible to engineer the fitting curve by designing a suitable ratio between $d_1$ and $d_2$, such as establishing a nearly linear relationship between $f_\mathrm{INZ}$ and $d_1$ within the range of $0.05\lambda_m<d_1<0.1\lambda_m$. This linear relationship could have potential applications in phase modulators, optical filters, and frequency combs.
	
	\begin{figure}[pt]
		\centering\includegraphics[width=4.5 in]{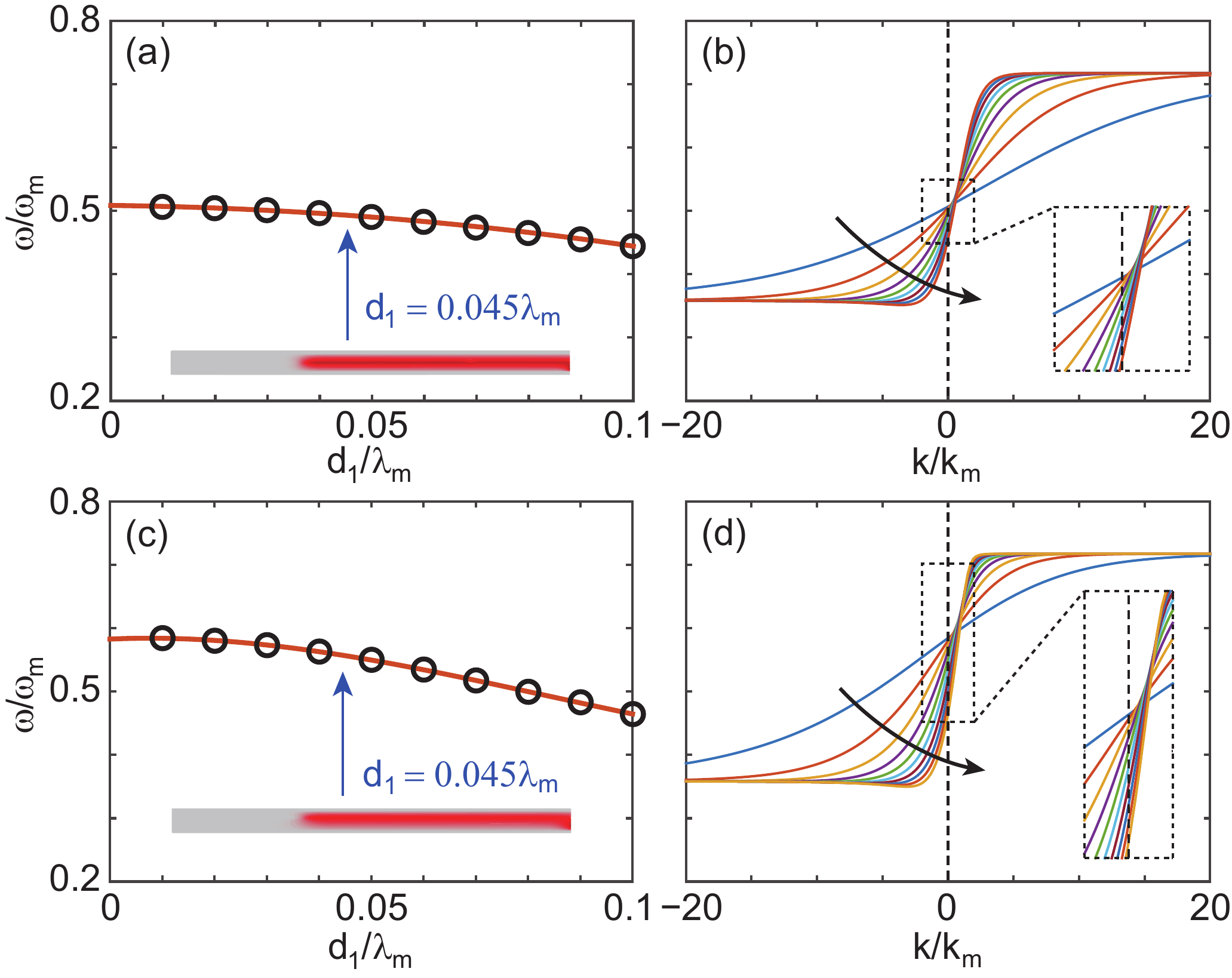}
		\caption{The operating frequencies of INZ modes as a function of $d_1$ in the cases of (a) $d_1=d_2$ and (c) $d_2=2 d_1$. The dispersion curves of one-way SMPs as the thicknesses changing (b) in case (a), and (d) in case (c).}\label{Fig3}
	\end{figure}
	\begin{figure}[ht]
		\centering\includegraphics[width=4.5 in]{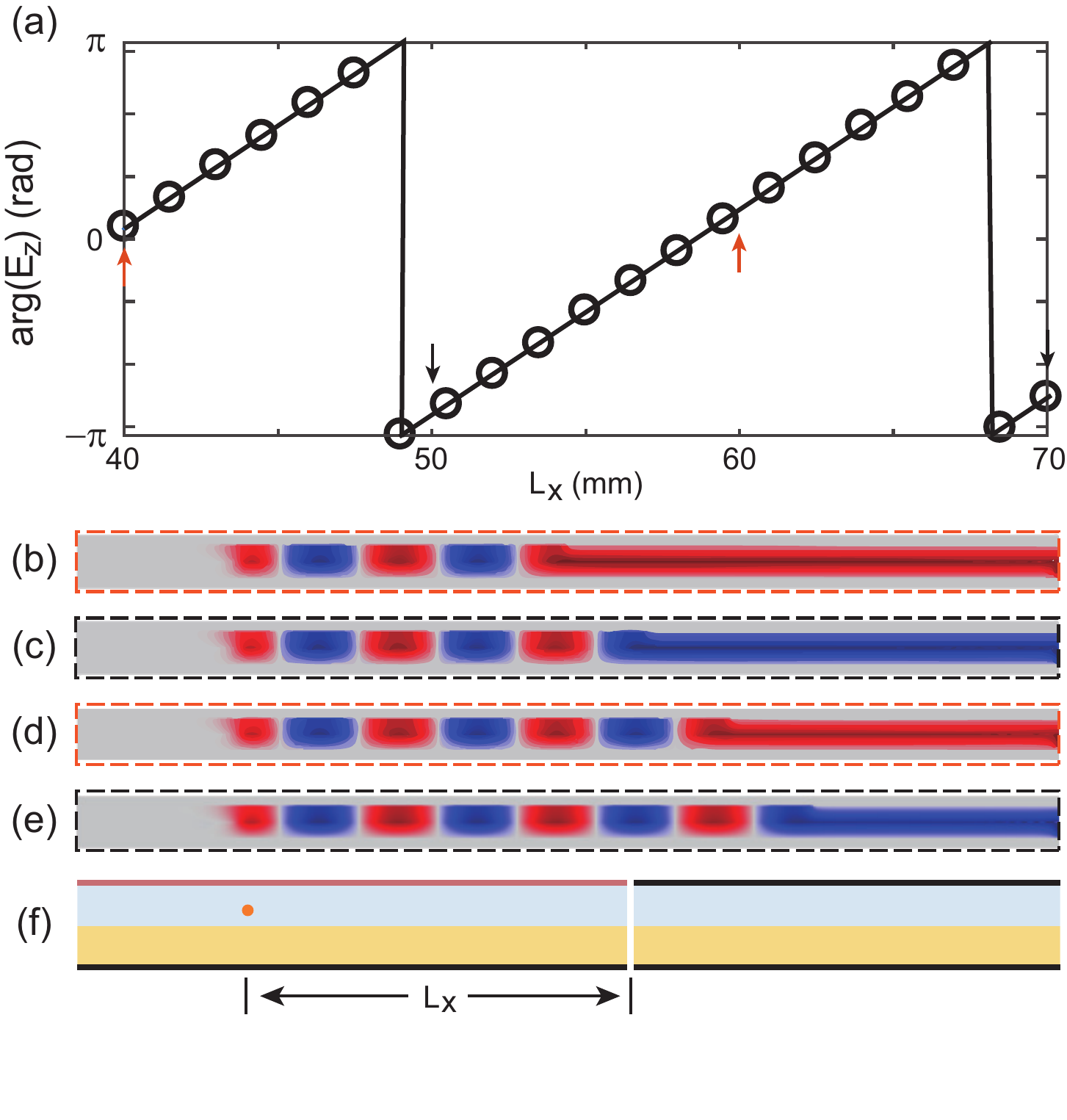}
		\caption{ All-optical phase modulator relied on INZ modes. (a) The phase angle of $E_z$ of the output INZ modes as a function of $L_x$. (b)-(e) Electric filed distributions of the FEM simulations, and $L_x$ are equal to (b) 40 mm, (c) 50 mm, (d) 60 mm, and (e) 70 mm, respectively. (f) The schematic picture of the INZ modes-based phase modulator. Parameters are $d_1=d_2=0.03\lambda_m$, $\nu=10^{-3}\omega_m$, and $f = 0.5007f_m$.}\label{Fig4}
	\end{figure}
	
	It is well known that an INZ wave can travel with no phase shift, which provides a potential approach to building a phase modulator. By designing a joint structure that contains two one-way waveguides with the same one-way region, and if the second waveguide supports the one-way INZ mode, it is theoretically possible to control the output phase by manipulating the joint position or the lengths of the waveguides. As analyzed in Fig. 2, the EDYE structure possesses the same one-way region as the MDYE structure, and an INZ mode is present in the EDYE structure. Therefore, as shown in Fig. 4(f), we propose a novel phase modulator called the 'MDYE-EDYE' structure. We further investigate how the joint position $L_x$ affects the output phase of the electric field ($E_z$). In Fig. 4(a), a clear linear relationship between the phase (arg($E_z$)) and $L_x$ is observed, suggesting that such a simple structure can achieve precise phase modulation.
	
	From Fig. 4(a), it is evident that to achieve a phase shift range of $[-\pi, \pi]$, the position shift $\Delta L_x$ should be approximately 20 mm, whereas the theoretical value in the lossless condition is around 19.3125 mm ($k \simeq 3.1068 k_m$). To clearly demonstrate the linear relationship between the phase angle of $E_z$ and $L_x$, we perform four full-wave simulations in a lossy condition with $\nu = 10^{-3}\omega_m$, where $L_x$ is set to 40 mm (Fig. 4(b)), 50 mm (Fig. 4(c)), 60 mm (Fig. 4(d)), and 70 mm (Fig. 4(e)). Consequently, the output $E_z$ in Figs. 3(b) and 3(d) are both positive with nearly the same phase, as well as in Figs. 3(c) and 3(e). Therefore, it is reasonable to believe that even in a real lossy condition, our proposed 'MDYE-EDYE' structure can function as a precise all-optical phase modulator, offering advantages such as robustness against backscattering reflection and a straightforward manufacturing process. Compared to previous all-optical phase modulators based on Mach-Zehnder Interferometers \cite{sturm:Alloptical,zhang:Ultralinear}, metasurfaces \cite{miao:Widely,yang:Selfbiased}, and/or metamaterials \cite{lee:Metamaterials}, our proposed structure exhibits comparable performance and simplicity.
	
	\section{Perfect filter, all-optical splitter, and self-consistent all-optical logic gates based on INZ modes}
	Furthermore, through careful assembly, the remanence-based MO system can function as a perfect filter, effectively separating electromagnetic (EM) waves within a specific frequency band into two separate channels with identical bandwidth. Fig. 5(b) illustrates the Y-shaped filter, which consists of three arms. The horizontal arm serves as the input port, while the other two arms, namely the 'EDYM' and 'EDYE' structures (previously studied in the previous subsection, Figs. 1(a) and 1(b)), function as the output ports. It is important to note that the magnetization direction in the 'EDYM' structure has been reversed to support the forward propagation of EM waves. The input port is referred to as the 'EYYE' structure, with the interconnected YIG layers having opposite magnetization directions. As shown in Fig. 5(a), the 'EYYE' structure exhibits a significantly wide one-way region, $0<\omega<\omega_r$, which corresponds to the combined one-way regions of the 'EDYE' structure ($0.5\omega_r<\omega<\omega_r$) and the 'EDYM' structure ($0<\omega<0.5\omega_r$).
	
	\begin{figure}[ht]
		\centering\includegraphics[width=4.5 in]{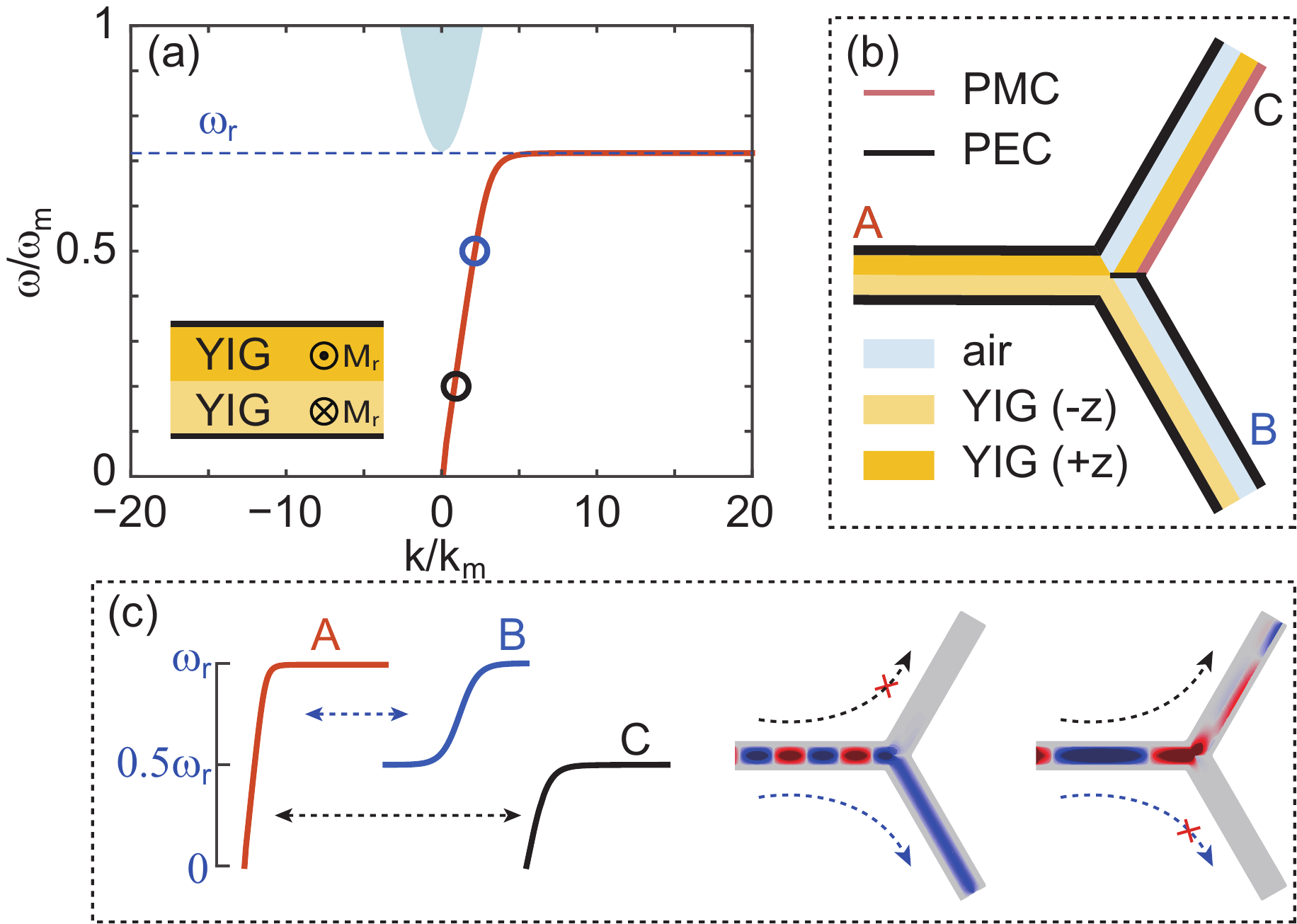}
		\caption{(a) Dispersion diagram of the YIG-YIG structure. (b) The schematic picture and (c) simulations of the perfect optical filter with frequencies are $f=0.5f_m$ (marked by blue circle in (a)) and $f=0.2f_m$ (marked by black circle in (a)), respectively.}\label{Fig5}
	\end{figure}
	
	Benefiting from the unique relationship between the one-way regions of the input port ('A') and the output ports ('B' and 'C'), along with the robustness of one-way EM waves, it can be theoretically guaranteed that the excited one-way EM wave from the input port will propagate into either of the output ports. To verify the performance of the filter, we conducted two finite element method (FEM) simulations in this structure with frequencies set at $f=0.5f_m$ (the second image in Fig. 5(c)) and $f=0.2f_m$ (the third image in Fig. 5(c)). As a result, the excited wave from 'A' propagated to the junction point and decisively injected into either 'B' or 'C', resulting in an extremely high contrast ratio between the output ports. Therefore, the simulation results align with our theoretical analysis. The proposed perfect filter has the potential to serve crucial roles such as an optical switch in an integrated optical circuit. By directly changing the working frequency, the state of the communication system can be switched between 'work' and 'stop'.
	
	\begin{figure}[ht]
		\centering\includegraphics[width=4.5 in]{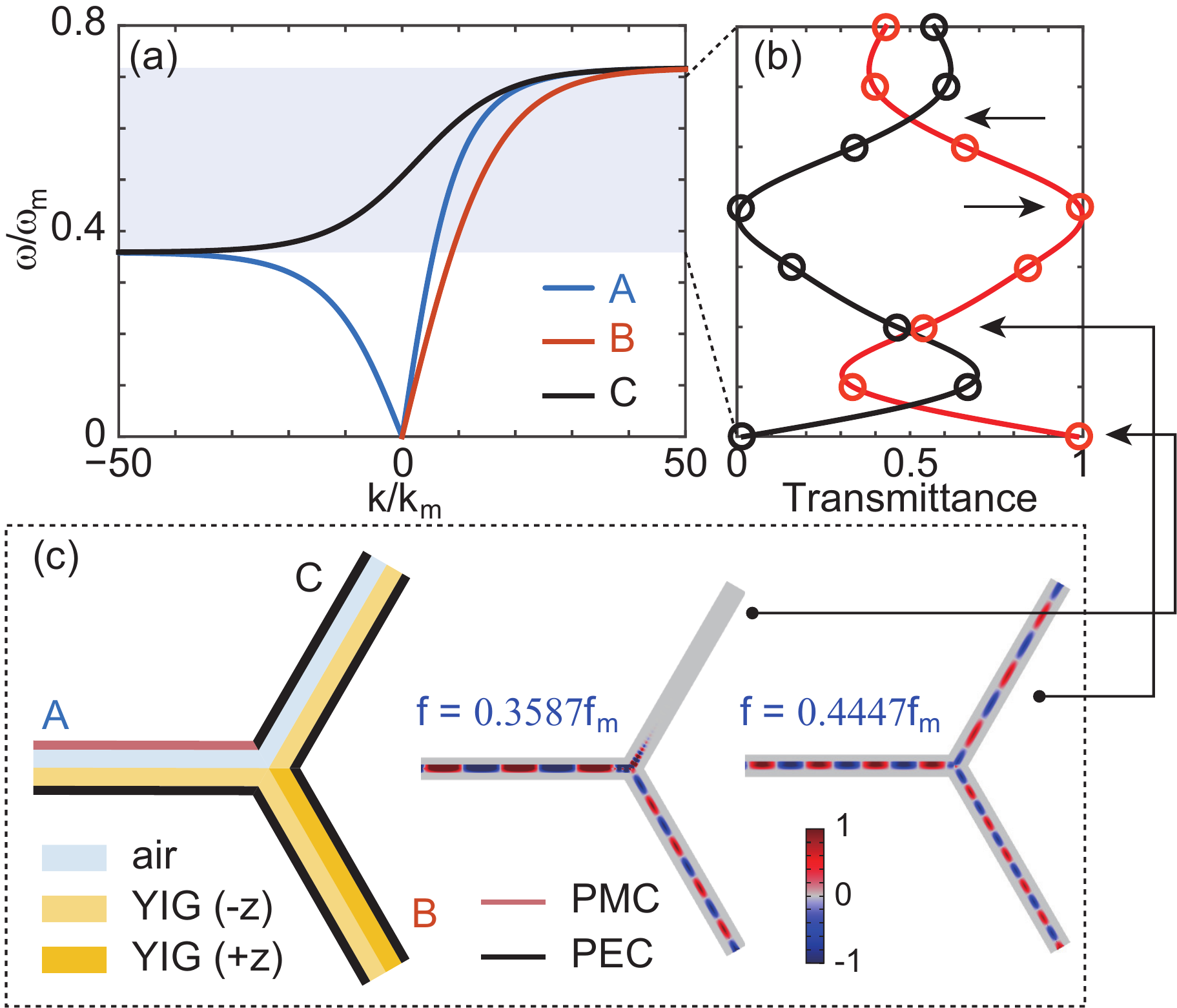}
		\caption{(a) Dispersion diagrams of three arms of a novel tunable splitter shown in (c). (b) Relationship between transmittances of two output arms and the operating frequency within the one-way region (colored area in (a)). (c) The schematic diagram and simulations for splitting ratio $\eta = 0/100$ (second picture) and $\eta = 50/50$ (last picture). Other parameters are $d_1=d_2=0.01\lambda_m$ and $\nu = 10^{-3}\omega_m$. }\label{Fig6}
	\end{figure}
	
	Besides, it is worth noting that the 'EDYE', 'MDYE', and 'EYYE' structures have overlapping one-way regions, but the wavenumbers of the one-way SMPs differ. As a result, one-way SMPs with different frequencies propagating in these structures will have different effective refractive indexes ($n_{eff}$) and may induce different splitting ratios ($\eta$). Fig. 6(c) illustrates the designed Y-shaped splitter, consisting of the 'MDYE' ('A'), 'EYYE' ('B'), and 'EDYE' ('C') structures. Figure 6(a) presents the dispersion curves of SMPs in the vicinity of the overlapping one-way region ($0.5\omega_r \leq \omega<\omega_r$). It is important to note that we set $d_1=d_2=0.01\lambda_m$ to engineer an SMP with an infinite wavenumber at $\omega=0.5\omega_r$ (the lower limit of the shaded region in Fig. 6(a)) in the 'EDYE' structure, which implies $n_{eff} \to \infty$ and consequently zero transmittance in the 'A-C' channel should be observed under this condition. We performed FEM simulations in the designed splitter, with the frequency changing within the one-way region, and calculated the transmittance. As depicted in Fig. 6(b), the red and black lines represent the transmittance of the 'A-B' and 'A-C' channels, respectively. As expected, when $\omega=0.5\omega_r=0.3587\omega_m$, nearly 100\% of the energy traveled through the 'A-B' channel (as shown in the second picture in Fig. 6(c)). More importantly, as the frequency gradually increases to $0.4447\omega_m$, the splitting ratio can approach 50/50 (as shown in the last picture in Fig. 6(c)), which is a highly desirable value in the field of optical communication. Another noteworthy case occurs when $\omega \simeq 0.45\omega_m$ (indicated by the right arrow in Fig. 6(b)). In this scenario, the EM wave cannot pass through the 'A-C' channel either. This behavior can be explained by the theory of total internal reflection: 1) the effective indices of 'A', 'B', and 'C' are approximately 19.4082 ($k\simeq10.6745k_m$), 29.4398 ($k\simeq16.1919 k_m$), and 6.8467 ($k\simeq3.7657k_m$), respectively; 2) the incident angle is $\pi/3$, which is clearly larger than the 'critical angle' (equal to $arcsin(6.8467/19.4082)$) of the 'A-C' channel. When the frequency is further increased to approximately $0.6456\omega_m$, a 50/50 splitting ratio will be achieved (indicated by the black left arrow in Fig. 6(b)). Therefore, our proposed remanence-based all-optical splitter has the capability to maintain a constant splitting ratio for different frequencies. For example, it can achieve $\eta=50/50$ at $\omega=0.4447\omega_m$ and/or $\omega=0.6456\omega_m$, which we believe is extremely important for parallel optical computing.
	
	\begin{figure}[ht]
		\centering\includegraphics[width=4.5 in]{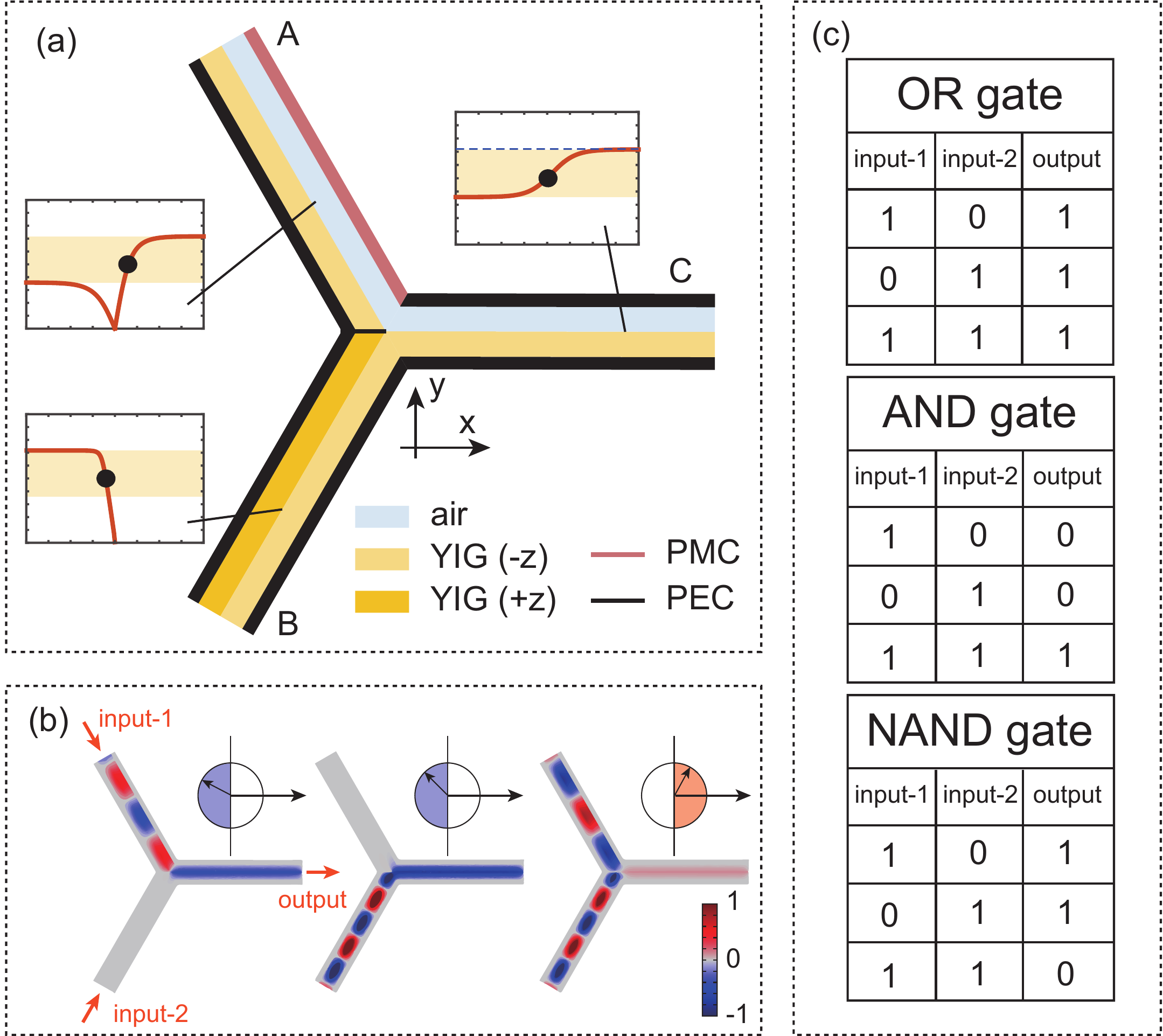}
		\caption{ (a) Schematic picture of all-optical LGs based on YIG with remanence. (b) Logic operations and FEM simulations for three different inputs, i.e. ('1', '0'), ('0', '1'), and ('1', '1'). (c) The truth table for OR, AND, and NAND gates according to simulation results shown in (b). }\label{Fig7}
	\end{figure}
	In our previous work\cite{Xu:All}, we proposed a possible method to achieve all-optical LGs with an extremely high contrast ratio. However, there was an unresolved issue in the previously proposed LGs, which involved inconsistency between the logic of the input and output ports. In this study, we propose another Y-shaped system based on INZ modes, which can function as all-optical LGs such as OR, AND, and NAND gates. Following a consistent positive logic convention, the presence of EM energy is treated as logic '1' and is utilized in both the input and output ports, while the phase angle of $E_z$ serves as the reference quantity in the output port. Figure 7(a) illustrates the schematic diagram of the designed system. One can observe that this structure is very similar to the one shown in Fig. 6(c), except for the exchange of YIG layers in arm 'B'. As discussed earlier, all three arms share the same one-way region, $0.5\omega_r<\omega<\omega_r$. In contrast to arm 'B' shown in Fig. 6, the arm in Fig. 7 can sustain backward-propagating SMPs (as shown in the inset of Fig. 7(a)). Consequently, two one-way channels are successfully engineered based on the splitter model shown in Fig. 6. In contrast to the working frequency used in our previous work, we choose the INZ frequency in the output port as the working frequency in order to preserve the phase information in the output section.
	
	The proposed system can be theoretically treated as a combination of basic LGs such as OR, AND, and NAND gates. For the OR gate, we utilize positive logic, where any presence of EM energy is considered logic '1'. Based on the robust one-way channels, any input EM signal (logic '1') from either arm 'A' or arm 'B' can unidirectionally propagate through arm 'C' (also logic '1'). Hence, the system naturally functions as an OR gate.
	
	To implement an AND gate, we introduce a reference quantity, the phase angle ($\theta$) of $E_z$, to determine the state of the output port. Specifically, we assume that the output EM signal with $\pi/2 < \theta < \pi/2$ represents logic '1'. As illustrated in Fig. 7(b), three input combinations, ['1', '0'], ['0', '1'], and ['1', '1'], yield three corresponding outputs, logic '0', logic '0', and logic '1'. It should be noted that we set $\theta \simeq 1.4$ rad for input-1 to engineer an appropriate output phase angle. Therefore, in practical applications, the input $\theta$ needs to be tailored for different inputs. In Fig. 7, the inputs $[\theta_1, \theta_2]$ are [-1.9 rad, ], [, 1.4 rad], and [1.4 rad, 1.4 rad], respectively.
	
	Figure 7(c) presents the truth table for the OR, AND, and NAND gates. It is evident that the outputs are inverted for the AND and NAND gates. Therefore, based on the analysis of the AND gate, the system can easily achieve NAND operation by assuming that the output EM signal with $\pi/2 < \theta < \pi/2$ represents logic '0'. Furthermore, other basic LGs can be straightforwardly extended. With improved integration and consistent logic, the proposed Y-shaped LGs enhance the potential applications of one-way mode-based LGs in the all-optical calculation of integrated optical circuits.
	
	\begin{figure}[ht]
		\centering\includegraphics[width=5 in]{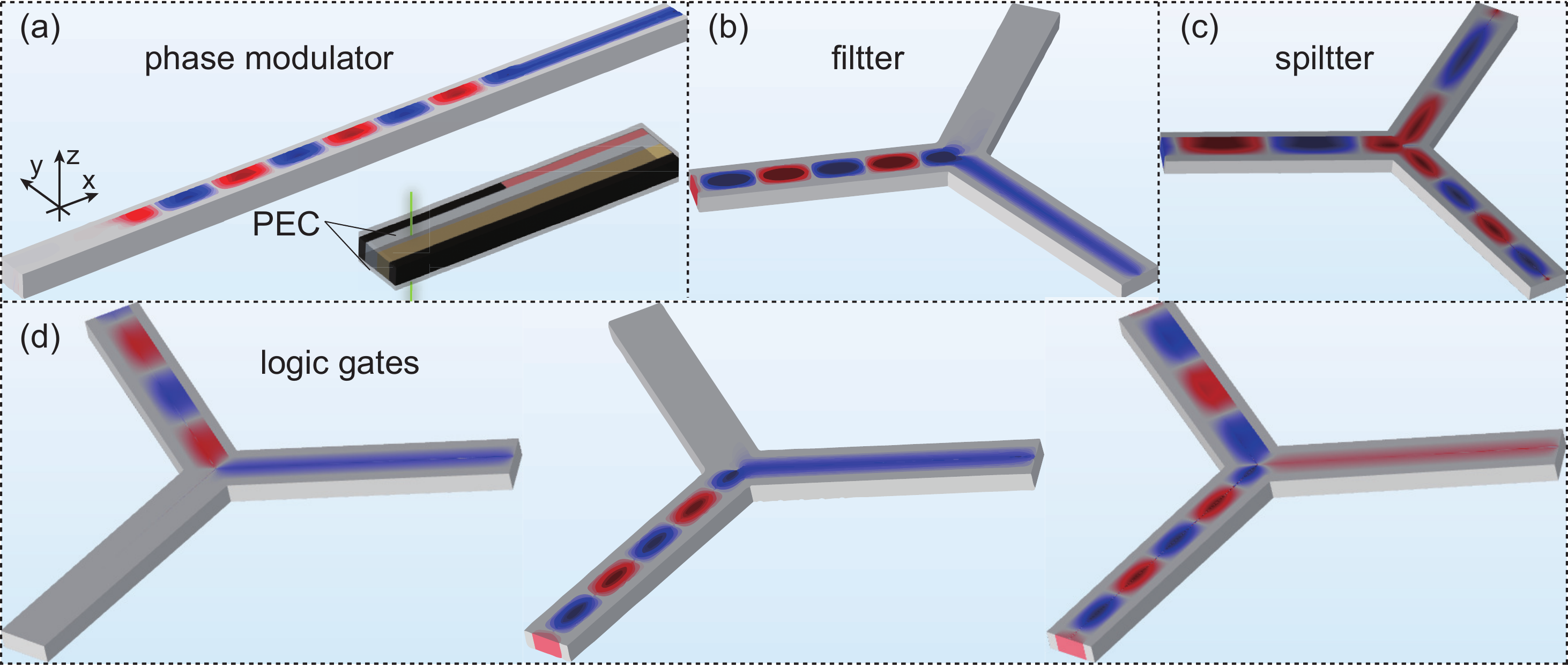}
		\caption{ Realization of 3D all-optical phase modulator, perfect filter, and self-consistent LGs.}\label{Fig8}
	\end{figure}

	The above-mentioned all-optical phase modulator, filter, and LGs, which are based on YIG with remanence, can be straightforwardly extended from 2D models to 3D structures\cite{Hong:Br}. As depicted in the inset of Fig. 8(a), two PEC walls are introduced in the z direction to confine the guided wave within the device. The simulation result shown in Fig. 8(a) exhibits similarities to the corresponding 2D case (Fig. 4(e)). Figures 7(b)-(e) illustrate the simulation outcomes for the 3D perfect filter, 3D splitter, and 3D LGs, respectively, and all the results are consistent with their 2D counterparts. Consequently, our proposed all-optical structures hold great potential for practical device implementation. Due to the characteristic of one-way propagation, the manufacturing process should be significantly simpler compared to other competing technologies. More importantly, as depicted in Fig. 1, our proposed all-optical phase modulator, filter, splitter, and LGs can be easily assembled on a chip. Therefore, it holds significant promise for enabling flexible all-optical computation and communication.

	\section{Conclusion}
	In conclusion, we have proposed a series of all-optical devices based on YIG with remanence. The all-optical phase modulator utilizes INZ modes and allows for controllable phase modulation in a linear relationship with the length of the boundary. The INZ frequency can be adjusted by manipulating the thickness parameters, providing tunability. Moreover, we have designed a perfect filter, which separates the input port's one-way region into two parts and creates two one-way channels with equal bandwidth. Additionally, we have introduced an all-optical splitter that divides the input wave into two output ports with a customizable and precise splitting ratio, such as 50/50. Furthermore, we have proposed all-optical logic gates based on one-way SMPs, where consistent positive logic and the phase angle of the output field are used to determine the logic states. Basic logic gates, including OR, AND, and NAND, have been achieved based on this principle. The feasibility and performance of the proposed devices have been validated in both 2D and 3D simulations. Our proposed all-optical devices, leveraging MO heterostructures, hold potential for optical calculations, including parallel computing in integrated optical circuits.
	
	\section*{Acknowledgement}
	This work was supported by the National Natural Science Foundation of Sichuan Province (No. 2023NSFSC1309), and the open fund of Luzhou Key Laboratory of Intelligent Control and Application of Electronic Devices (No. ZK202210), Sichuan Science and Technology Program (No. 2022YFS0616), the Science and Technology Strategic Cooperation Programs of Luzhou Municipal People’s Government and Southwest Medical University (No. 2019LZXNYDJ18). J.X. and Y.L. thanks for the support of the Innovation Laboratory of Advanced Medical Material \& Physical Diagnosis and Treatment Technology. K.L.T. was supported by the General Secretariat for Research and Technology (GSRT) and the Hellenic Foundation for Research and Innovation (HFRI) under Grant No. 4509.

	
	\bibliography{mybib}
	

\end{document}